\documentclass[preprint,12pt]{elsarticle}

\usepackage{amssymb}
\usepackage{amsmath}

\usepackage{color}

\journal{Computer Physics Communications}

\begin{document}

\begin{frontmatter}



\title{\emph{Menura}: a code for simulating the interaction between a turbulent solar wind and solar system bodies}


\author[inst1,inst2]{E. Behar}
\author[inst3]{S. Fatemi}
\author[inst2,inst4]{P. Henri}
\author[inst1]{M. Holmstr\"om}

\affiliation[inst1]{organization={Swedish Institute of Space Physics},
            city={Kiruna},
            country={Sweden}}
\affiliation[inst2]{organization={Laboratoire Lagrange, Observatoire de la C\^ote d'Azur, Universit\'e C\^ote d'Azur, CNRS},
            city={Nice},
            country={France}}
\affiliation[inst3]{organization={Department of Physics at Ume\aa \ University, Ume\aa, Sweden}}
\affiliation[inst4]{organization={LPC2E},
            city={Orl\'eans},
            country={France}}



\begin{abstract}
Despite the close relationship between planetary science and plasma physics, few advanced numerical tools allow to bridge the two topics. The code \emph{Menura} proposes a breakthrough towards the self-consistent modelling of these overlapping field, in a novel 2-step approach allowing for the global simulation of the interaction between a fully turbulent solar wind and various bodies of the solar system. This article introduces the new code and its 2-step global algorithm, illustrated by a first example: the interaction between a turbulent solar wind and a comet.
\end{abstract}



\begin{keyword}
Solar Wind \sep Turbulence \sep Magnetosphere \sep Hybrid Simulation
\PACS 0000 \sep 1111
\MSC 0000 \sep 1111
\end{keyword}

\end{frontmatter}


\section{Introduction}
\label{sec:intro}

    For about a century, three main research fields have taken an interest in the various space plasma environments found around the Sun. On the one hand, two of them, namely planetary science and solar physics, have been exploring the solar system, to understand the functioning and history of its central star, and of its myriad of orbiting bodies. On the other hand, the third one, namely fundamental plasma physics, has been using the solar wind as a handy wind tunnel which allows researchers to study fundamental plasma phenomena not easily reproducible on the ground in laboratories. During the last two decades, bridges between these communities have been developing, as the growing knowledge of each community was bringing the fields ever closer, to a point where overlapping topics made the communities work together. For instance, if initially planetary scientists were studying the interaction between solar system bodies and a steady, ideally laminar solar wind, they soon had to consider the event-full and turbulent nature of the solar wind to go further in the in situ space data analysis, further in their understanding of the interactions at various obstacles. If plasma physicists were originally interested in a pristine solar wind unaffected by the presence of obstacles, they realised that the environment close to these obstacles could provide combinations of plasma parameters otherwise not accessible to their measurements in the unaffected solar wind. For a while now, we have seen planetary studies focusing on the effects of solar wind transient effects (such as Coronal mass Ejection CME or Co-rotational Interaction Region CIR) on planetary plasma environments, at Mars \cite{ramstad2017grl}, Mercury \cite{exner2018pss}, Venus \cite{luhmann2008jgr} and comet 67P/C-G \cite{edberg2016mnras, hajra2018mnras} to only cite a few, the effect of large scale fluctuations in the upstream flow on Earth's magnetosphere \cite{tsurutani1987pss}, and more generally the effect of solar wind turbulence on Earth magnetosphere and ionosphere \cite{damicis2020fass, guio2021fass}. Similarly, plasma physicists have developed comprehensive knowledge of plasma waves and plasma turbulence in the Earth magnetosheath, presenting relatively high particle densities and electromagnetic field strengths favourable for space instrumentation, in a region more easily accessible to space probes than regions of unaffected solar wind \cite{borovsky2003jgr, rakhmanova2021fass}. More recently, the same community took an interest in various planet magnetospheres, depicting plasma turbulence in various locations and of various parameters \cite{saur2021fass} and all references therein.
    
    Various numerical codes have been used for the global simulation of the interaction between a laminar solar wind and solar system bodies, using MHD \cite{gombosi2004cse}, hybrid \cite{bagdonat2002jcp}, or fully kinetic \cite{markidis2010mcs} solvers. Similarly, solar wind turbulence in the absence of an obstacle has also been simulated using similar MHD \cite{boldyrev2011apj}, hybrid \cite{franci2015apj}, and fully kinetic \cite{valentini2007jcp} solvers. In this context, we identify the lack of a numerical approach for the study of the interaction between a turbulent plasma flow (such as the solar wind) and an obstacle (such as a magnetosphere, either intrinsic or induced). Such a tool would provide the first global picture of these complex interactions. By shading new lights on the long-lasting dilemma between intrinsic phenomena and phenomena originating from the upstream flow, it would allow invaluable comparisons between self-consistent, global, numerical results, and the worth of observational results provided by the various past, current and future exploratory space missions in our solar system.\\
    
    The main points of interest and main questions motivating such a model can be organised as such:
    
    \begin{itemize}
        \item Macroscopic effects of turbulence on the obstacle
        \begin{itemize}
            \item shape and position of the plasma boundaries (e.g. bow shock, magnetopause),
            \item large scale magnetic reconnection,
            \item atmospheric escape,
            \item dynamical evolution of the magnestosphere.
        \end{itemize}
        \item Microscopic physics and instabilities within the interaction region, induced by upstream turbulence
        \begin{itemize}
            \item energy transport by plasma waves,
            \item energy conversion by wave-particle interactions,
            \item energy transfers by instabilities.
        \end{itemize}
        \item The way incoming turbulence is processed by planetary plasma boundaries
        \begin{itemize}
            \item sudden change of spatial and temporal scales,
            \item change of spectral properties,
            \item existence of a memory of turbulence downstream magnetospheric boundaries.
        \end{itemize}
    \end{itemize}
    
    Indirectly, because of the high numerical resolution required to properly simulate plasma turbulence, this numerical experiment will provide an exploration of the various obstacles with the same high resolution in both turbulent and laminar runs, resolutions that have rarely been reached for planetary simulations, except for Earth's magnetosphere.\\
    
    \emph{Menura}, the new code presented in this publication, splits the numerical modelling of the interaction into two steps. Step 1 is a decaying turbulence simulation, in which electromagnetic energies initially injected at the large spatial scales of the simulation box cascades towards smaller scales. Step 2 uses the output of Step 1 to introduce an obstacle moving through this turbulent solar wind.
    
    The code is written in \texttt{c++} and uses \texttt{CUDA} APIs for running its solver exclusively on Graphics Processing Units (GPUs). Section \ref{sec:solver} introduces the solver, which is tested against classical plasma phenomena in Section \ref{sec:physical_tests}. Sections \ref{sec:step1} and \ref{sec:step2} tackle the first and second step of the new numerical modelling approach, illustrating the decaying turbulence phase, and introducing the algorithm for combining the output of Step 1 together with the modelling of an obstacle (Step 2). Section \ref{sec:result} presents the first global result of \emph{Menura}, providing a glimpse of the potential of this numerical approach, and introducing the forthcoming studies.\\
    
    \emph{Menura} source code is open source, available under the GNU General Public License license.

\section{The solver}
\label{sec:solver}

    In order to (i) achieve global simulations of the interactions while (ii) modelling the plasma kinetic behaviour, with regard to the computation capabilities currently available, a hybrid Particle-In-Cell (PIC) solver has been chosen for \emph{Menura}. This well-established type of models resolves the Vlasov equation for the ions by discretising the ion distribution function as macro-particles characterized by discrete positions in phase space, and electrons as a fluid, with characteristics evaluated at the nodes\footnote{Only the discrete nodes of the grid are considered in the solver, though the term ``cell'' is equivalently used by other authors.} of a grid, together with ion moments and electromagnetic fields. The fundamental computational steps of a hybrid PIC solver are:
    
    \begin{itemize}
        \item Particles' position advancement, or ``push''.
        \item Particles' moments mapping, or ``gathering'': density, current, eventually higher order, as required by the chosen Ohm's law.
        \item Electromagnetic field advancement, using either an ideal, resistive or generalised Ohm's law and Faraday's law.
        \item Particles' velocity advancement, or ``push''.
    \end{itemize}
    
\begin{figure}
        \centering
        \includegraphics[width=\textwidth]{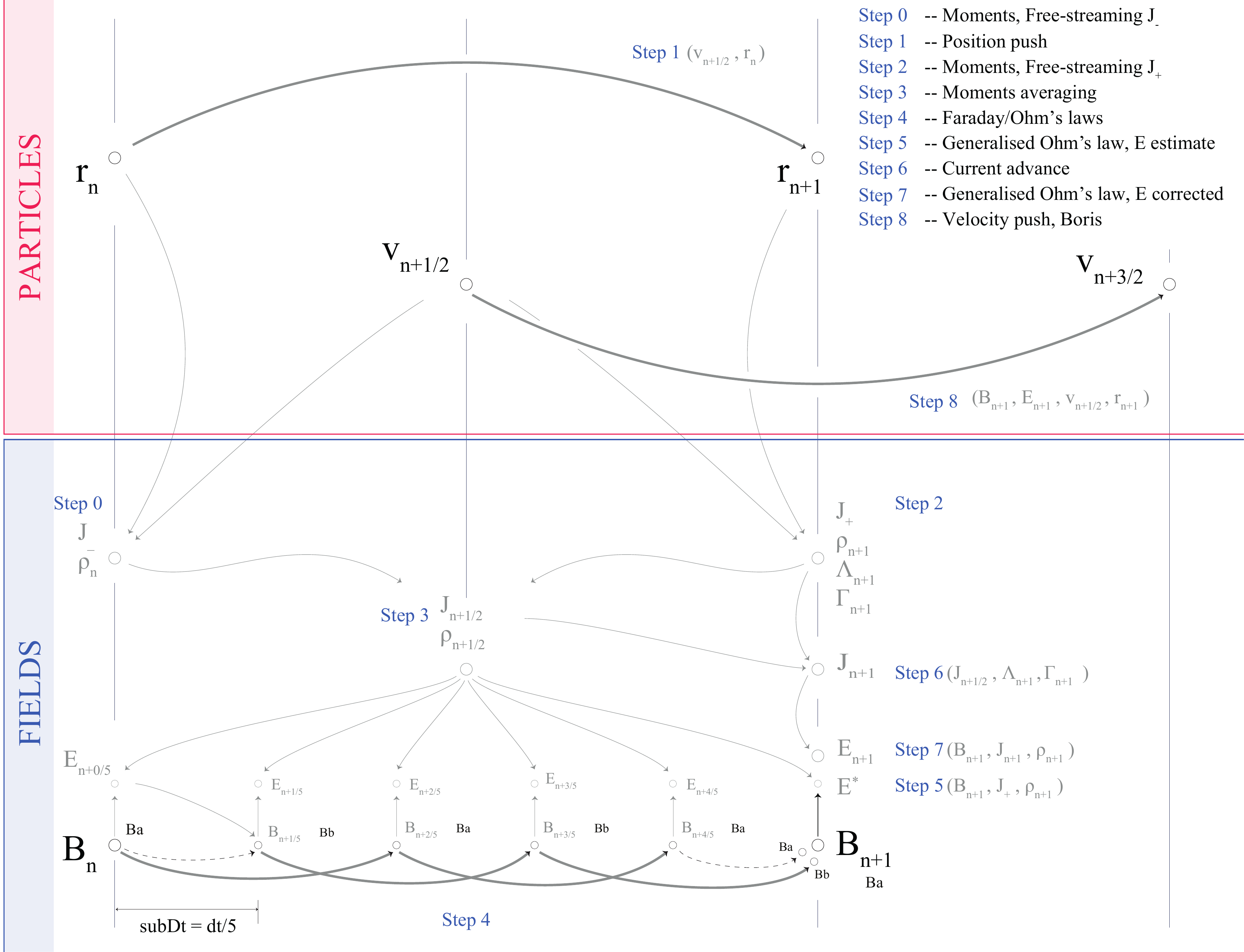}
        \caption{Algorithm of \emph{Menura}'s solver, with its main operations numbered from 0 to 8, as organised in the \texttt{main} file of the code. $\mathbf{r}$ and $\mathbf{v}$ are the position and velocity vectors of the macro particles. Together with the magnetic field $\mathbf{B}$, they are the only variables necessary for the time advancement. The electric field $\mathbf{E}$, the current $\mathbf{J}$, the charge density $\rho$, as well as the CAM pseudo-moments $\Lambda$ and $\Gamma$, are obtained from $\mathbf{r}$, $\mathbf{v}$ and $\mathbf{B}$.}
        \label{fig:algo}
    \end{figure}

    These steps are summarised in Figure \ref{fig:algo_inj}. Details about these classical principles can be found in \cite{tskhakaya2008} and references therein. The bottleneck of PIC solvers is the particles' treatment, especially the velocity advancement and the moments computation (namely density and current). The simulation of plasma turbulence especially requires large amounts of macro-particles per grid nodes. We therefore want to minimise both the amount of operations done on the particles and the number of particles itself. A popular method which minimises the amount of these computational passes through all particles is the Current Advance Method (CAM) \cite{matthwes1994jcp}, for instance used for the hybrid modelling of turbulence by \cite{franci2015apj}. Figure \ref{fig:algo} presents \emph{Menura}'s solver algorithm, built around the CAM, similar to the implementation of \cite{bagdonat2002jcp}. In this scheme, only four passes through all particles are performed, one position and one velocity pushes and two particle moments mappings. The second moment mapping in Figure \ref{fig:algo}, i.e. step 2, also produces the two pseudo-moments \texttt{$\Lambda$} and \texttt{$\Gamma$} used to advance the current as:
    
    \begin{align}
        \Lambda & = \sum_{p} \frac{q^2}{m} W(\mathbf{r}_{n+1}) , \\
        \Gamma & = \sum_{p} \frac{q^2}{m} \mathbf{v}_{n+1/2} W(\mathbf{r}_{n+1}) , \\
        J_{n+1} & = J_{n+1/2} + \frac{\Delta t}{2} (\Lambda \mathbf{E}^* + \Gamma \times \mathbf{B}) ,
    \end{align}
    
    with $\mathbf{E}^*$ the estimated electric field after the magnetic field advancement of step 4. $W(\mathbf{r}_{n+1})$ is the shape function, which attributes different weights for each node surrounding the macro-particle \cite{tskhakaya2008}.
    
    Central finite differences for evaluating derivatives and second order interpolations are used throughout the solver. The grid covering the physical simulation domain has an additional 2-node wide band, the guard or ghost nodes, allowing to solve derivatives using (central) finite differences at the very edge of the physical domain. For periodic boundary conditions, as used along all directions during Step 1 of the simulation, the value at the opposite edge of the physical domain are copied in the guard nodes. Other boundary conditions will be discussed later when introduced.
    
    The mapping of the particle moments are done using an order-two, triangular shape function: one macro-particle contributes to 9 grid nodes in 2D space ( respectively 27 in 3D space), using 9 (respectively 27) different weights. The interpolation of the fields values from the nodes to the macro-particles' positions uses the exact same weights, with 9 (respectively 27) neighbouring nodes contributing to the fields values at a particle position.
    
    As illustrated in Figure \ref{fig:algo}, the position and velocity advancements are done at interleaved times, similarly as a classical second order leap-frog scheme. However, since the positions of the particles are needed to evaluate their acceleration, the CAM scheme is not strictly speaking a leap-frog integration scheme. Another difference in this implementation is that velocities are advanced using the Boris method \cite{boris1970relativistic}.\\
    
    The Ohm's law is at the heart of the hybrid modelling of plasmas. \emph{Menura} uses the following form of the law, here given in SI units. In this formulation, the electron inertia is neglected, and the quasi-neutral approximation $n\sim n_i \sim n_e$ is used \cite{valentini2007jcp}. Additionally, neglecting the time derivative of the electric field in the Ampere-Maxwell's law (Darwin's hypothesis), one get the total current through the curl of the magnetic field. This formulation highlights the need for only three variables to be followed through time, namely the magnetic field, and the particles position and velocity, while all other variables can be reconstructed from these three.
    
    \begin{equation}\label{eq:ohm}
        \mathbf{E} = -\mathbf{u_i}\times\mathbf{B} + \frac{1}{e\, n \mu_0}  \mathbf{J} \times \mathbf{B} + \frac{1}{e\, n} \nabla\cdot p_e - \eta_h \nabla^2 \mathbf{J}
    \end{equation}
    
    The Faraday's law is used for advancing the magnetic field in time:
    
    \begin{equation}
        \frac{\partial \mathbf{B}}{\partial t} = -\mathbf{\nabla}\times \mathbf{E}
    \end{equation}
    
    The electron pressure is obtained assuming it results from a polytropic process, with an arbitrary index $\kappa$, to be carefully chosen by the user.
    
    \begin{equation}
        p_e = p_{e0}\left(\frac{n_e}{n_{e0}}\right)^\kappa
    \end{equation}
    
    Using much less memory than the particles' variables, the fields can be advanced in time using a smaller time step and another leap-frog-like approach, as illustrated in Figure \ref{fig:algo}, step 4 \cite{matthwes1994jcp}.\\
    
    Additional spurious high-frequency oscillations are the default behaviour of finite differences schemes. Two main families of methods are used to filter out these features, the first being an additional step of field smoothing, the second using the direct inclusion of a diffusive term in the differential equation of the system, acting as a filter \cite{maron2008apj}. For \emph{Menura}, we have retained the second approach, implementing a term of hyper-resistivity in the Ohm's law.\\
    
    The stability of hybrid solvers is sensitive to low ion densities. We use a threshold value equal to a few percent of the background density, 5\% in the following examples, threshold below which a node is considered as a vacuum node, and only the resistive terms of the generalised Ohm's law of Equation \ref{eq:ohm} are solved using a higher value of resistivity $\eta_{h\  \text{vacuum}}$ \cite{holmstrom2013}. This way, terms proportional to $1/n$ do not exhibit nonphysical values where the density may get locally very low, due to the thermal noise of the PIC macro-particle discretisation.
    
    All variables in the code are normalised using the background magnetic field amplitude $B_0$ and the background plasma density $n_0$. All variables are then expressed in terms of either these two background values, or equivalently in terms of the proton gyrofrequency $\omega_{ci0}$ and the Alfven velocity $v_{A0}$. We define normalised variables $\tilde{a}$ as obtained by dividing its physical value by its ``background'' value:
    
    \begin{equation}
        \tilde{a} = \frac{a}{a_0}
    \end{equation}
    
    All background values are given in Table \ref{tab:normalisation}, and the normalised equations of the solver are given in \ref{app:normalised_equations}.
    
    \begin{table}[]
        \centering
        \begin{tabular}{c|c}
            $B_0$ & $B_0$ \\
            $n_0$ & $n_0$ \\
            $v_0$ & $v_{A0}=B_0/\sqrt{\mu_0 m_i n_0}$\\
            $\omega_0$ & $\omega_{ci0}=e B_0/m_i$\\
            $x_0$ & $d_{i0}=v_{A0}/\omega_{ci0}$ \\
            $t_0$ & $1/\omega_{ci0}$\\
            $p_0$ & $B_0^2/(2\mu_0)$\\
            $m_0$ & $m_i n_0 x_0^3$\\
            $q_0$ & $e\, n_0 x_0^3$\\
        \end{tabular}
        \caption{Caption}
        \label{tab:normalisation}
    \end{table}

\section{Physical tests}
\label{sec:physical_tests}

\begin{figure}
        \centering
        \includegraphics[width=\textwidth]{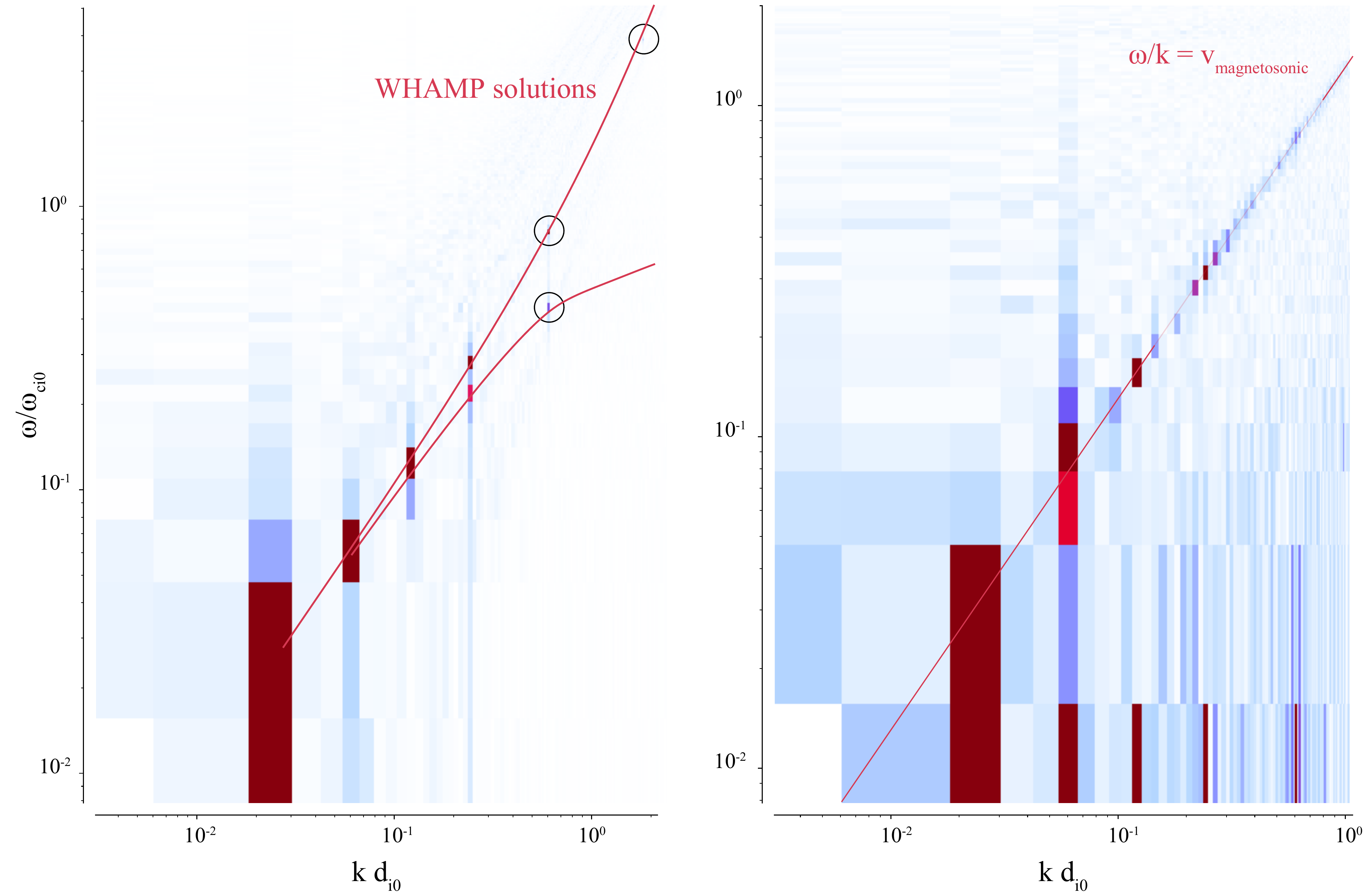}
        \caption{MHD modes dispersion relations, as solved by \emph{WHAMP} and \emph{Menura}. $B_0=1.8$ nT, $n_0=1.$ cm$^{-3}$, $T_{i0}=10^4$ K, $T_{e0}=10^5$ K}
        \label{fig:MHD}
    \end{figure}

    In this section, the code is tested against well-known, collisionless plasma processes, and their solutions given by the linear full kinetic solver \emph{WHAMP} \cite{ronnmark1982waves}. We first explore MHD scales, simulating Alfv\'enic and magnetosonic modes. We use a 2-dimensional spatial domain with one preferential dimension chosen as $\mathbf{x}$. A sum of six cosine modes in the component of the magnetic field along the $x$-direction direction are initialised, corresponding to the first six harmonics of this periodic box. The amplitude of these modes is 0.05 times the background magnetic field $B_0$, which is taken either along (Alfv\'en mode) or across (magnetosonic mode) the propagation direction $\mathbf{x}$. Data are recorded along time and along the main spatial dimension $\mathbf{x}$, resulting in the 2D field $B(x, t)$. The 2-dimensional Fourier transform of this field is given in Figure \ref{fig:MHD}. In this ($\omega, k$)-plane, each mode can be identified as a point of higher power, six points for six initial modes. The solutions given by \emph{WHAMP} for the same plasma parameters are shown by the solid lines, and a perfect match is found between the two models. Close to the ion scale $k.d_i=1$, \emph{WHAMP} and \emph{Menura} display two different branches that originate from the Alfv\'en mode, splitting for higher frequencies into the whistler and the ion cyclotron branches. The magnetosonic modes were also tested using different polytropic indices, resulting in a shift of the dispersion relation along the $\omega$-axis. Changing polytropic indices for both models resulted in the same agreement.
    
\begin{figure}
        \centering
        \includegraphics[width=\textwidth]{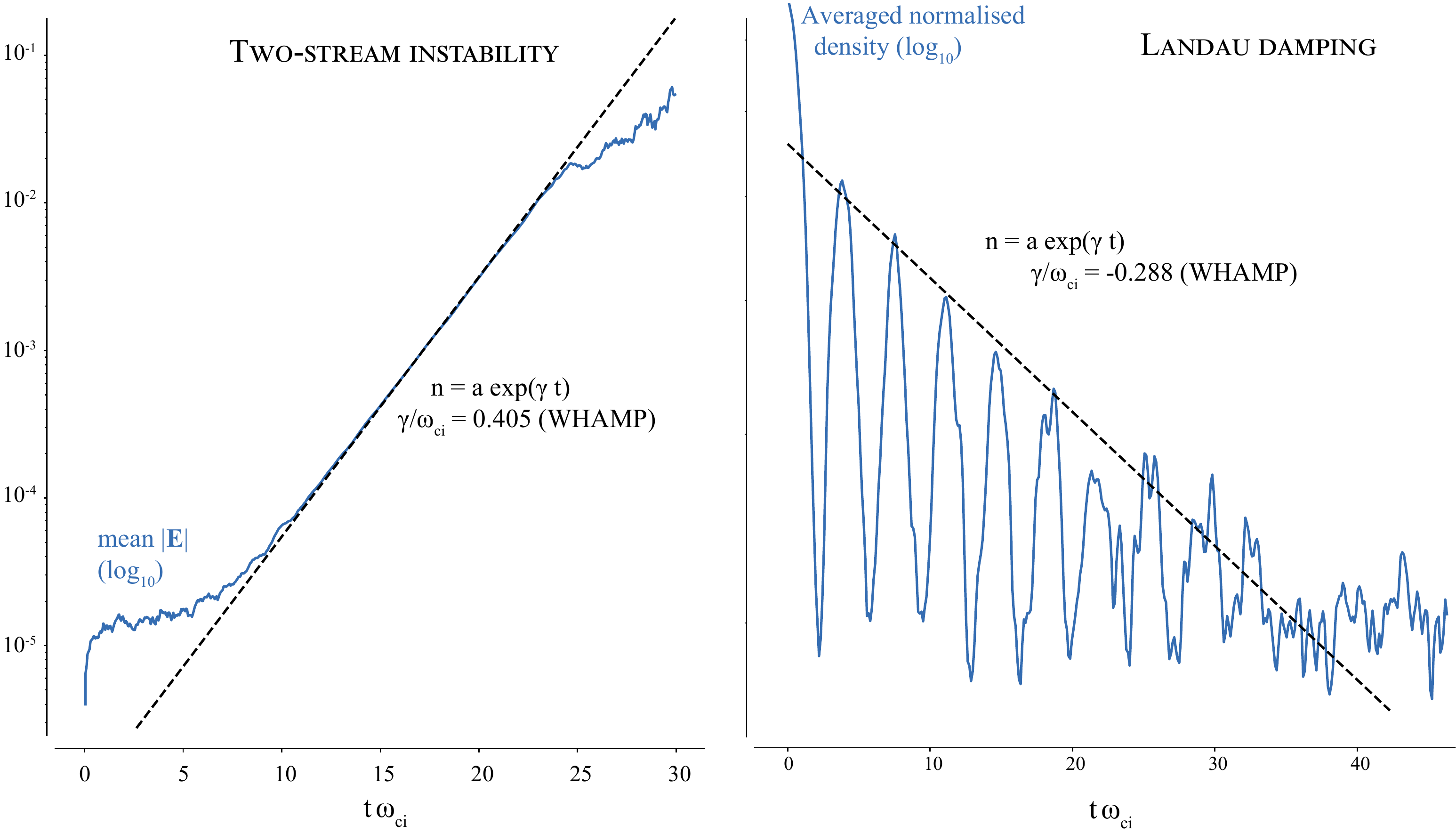}
        \caption{Left-hand side, growth during the linear phase of the ion-ion two-stream instability; Right-hand side, Landau damping of an ion acoustic mode. Two-stream instability: $B_0=1.8$ nT, $n_0=1.$ cm$^{-3}$, $T_{i0}=10^2$ K, $T_{e0}=10^3$ K. Landau Damping: $B_0=1.8$ nT, $n_0=5.$ cm$^{-3}$, $T_{i0}=1.5\cdot 10^4$ K, $T_{e0}=10^5$ K}
        \label{fig:kinetics}
    \end{figure}

    With the MHD scales down to ion inertial scales now validated, we explore the ability of the solver to account for further ion kinetic phenomena, first with the classical case of the two-stream instability (also known as the ion-beam instability, given the following configuration). Two Maxwellian ion beams are initialised propagating with opposite velocities along the main dimension $\mathbf{x}$. A velocity separation of $15 v_{th}$ is used in order to excite only one unstable mode. The linear kinetic solver \emph{WHAMP} is used to identify the expected growth rate associated to the linear phase of the instability, before both beams get strongly distorted and mixed in phase space during the nonlinear phase of the instability (not capture by WHAMP). During this linear phase, \emph{Menura} results in a growing circularly polarised wave, and the amplitude's growth of the wave is shown in Figure \ref{fig:kinetics}. Both growths match perfectly.
    
    Finally, we push the capacities of the model to the case of the damping of an ion acoustic wave through Landau resonance. A very high number of macro-particles per grid node is required to resolve this phenomenon, so enough resonant particles take part in the interaction with the wave. The amplitude of the initial, single acoustic mode is taken as 0.01 times the background density, taken along the main spatial dimension of the box. The decrease in the density fluctuation through time, spatially averaged, is shown in Figure \ref{fig:kinetics}, with again the corresponding solution from \emph{WHAMP}. A satisfying agreement is found during the first 6 oscillations, before the noise in the hybrid solver output (likely associated to the macroparticle thermal noise) takes over. Admittedly, the amount of particle per node necessary to well resolve this phenomenon is not practical for the global simulations which \emph{Menura} (together with all global PIC simulations) aims for.
    
    For the classical tests presented above, spanning over MHD and ion kinetic scales tests, \emph{Menura} agrees with theoretical and linear results. In the next section, the simulation of a decaying turbulent cascade provides one final physical validation of the solver, through all these scales at once.

\section{Step 1: Decaying turbulence}
    \label{sec:step1}
    
    We use \emph{Menura} to simulate plasma turbulence using a decaying turbulent cascade approach: at initial time $t=0$, a sum of sine modes with various wave vectors $\mathbf{k}$, spanning over the largest spatial scales of the simulation domain, are added to both the homogeneous background magnetic field $\mathbf{B}_0$ and the particles velocities $\mathbf{u}_i$. Particle velocities are initialised according to a Maxwellian distribution, with a thermal speed equal to one Alfv\'en speed. Without any other forcing later-on, this initial energy cascades, as time advances, towards lower spatial and temporal scales, while forming vortices and reconnecting current sheets \cite{franci2015apj}. Using such Alfv\'enic perturbation is motivated by the predominantly Alfv\'enic nature of the solar wind turbulence measured at 1 au \cite{bruno2013turbulence}.
    
    In this 2-dimensional set-up, $\mathbf{B}_0$ is taken along the $\mathbf{z}$-direction, perpendicular to the simulated spatial domain $(\mathbf{x}, \mathbf{y})$, whereas all initial perturbations are defined within the simulation plane. Their amplitude is 0.5 $B_0$, while their wave vectors are taken with amplitude between $k_\text{inj\, min}=0.01\ d_{i0}$ and $k_\text{inj\, max}=0.1\ d_{i0}$, so energy is only injected in MHD scales, in the inertial range. Because we need these perturbation fields to be periodic along both directions, the $k_x$ and $k_y$ of each mode corresponds to harmonics of the simulation box dimensions. Therefore, a finite number of wave vector directions is initialised. Along these constrained directions, each mode in both fields has two different, random phases. The magnetic field is initialised such that is it divergence-free.
    
    For this example, the box is chosen to be 500 $d_{i0}$ wide on both dimensions, subdivided by a grid 1000$^2$ nodes wide. The corresponding $\Delta x$ is 0.5 $d_{i0}$, and spatial frequencies are resolved over the range [0.0062, 6.2] $d_{i0}$. The time step is 0.05 $\omega_{ci0}^{-1}$. 2000 particles per grid node are initialised with a thermal speed of 1 $v_A$. The temperature is isotropic and a plasma beta of 1 is chosen for both the ion macro-particles and the electronic massless fluid.
    
\begin{figure}
        \centering
        \includegraphics[width=\textwidth]{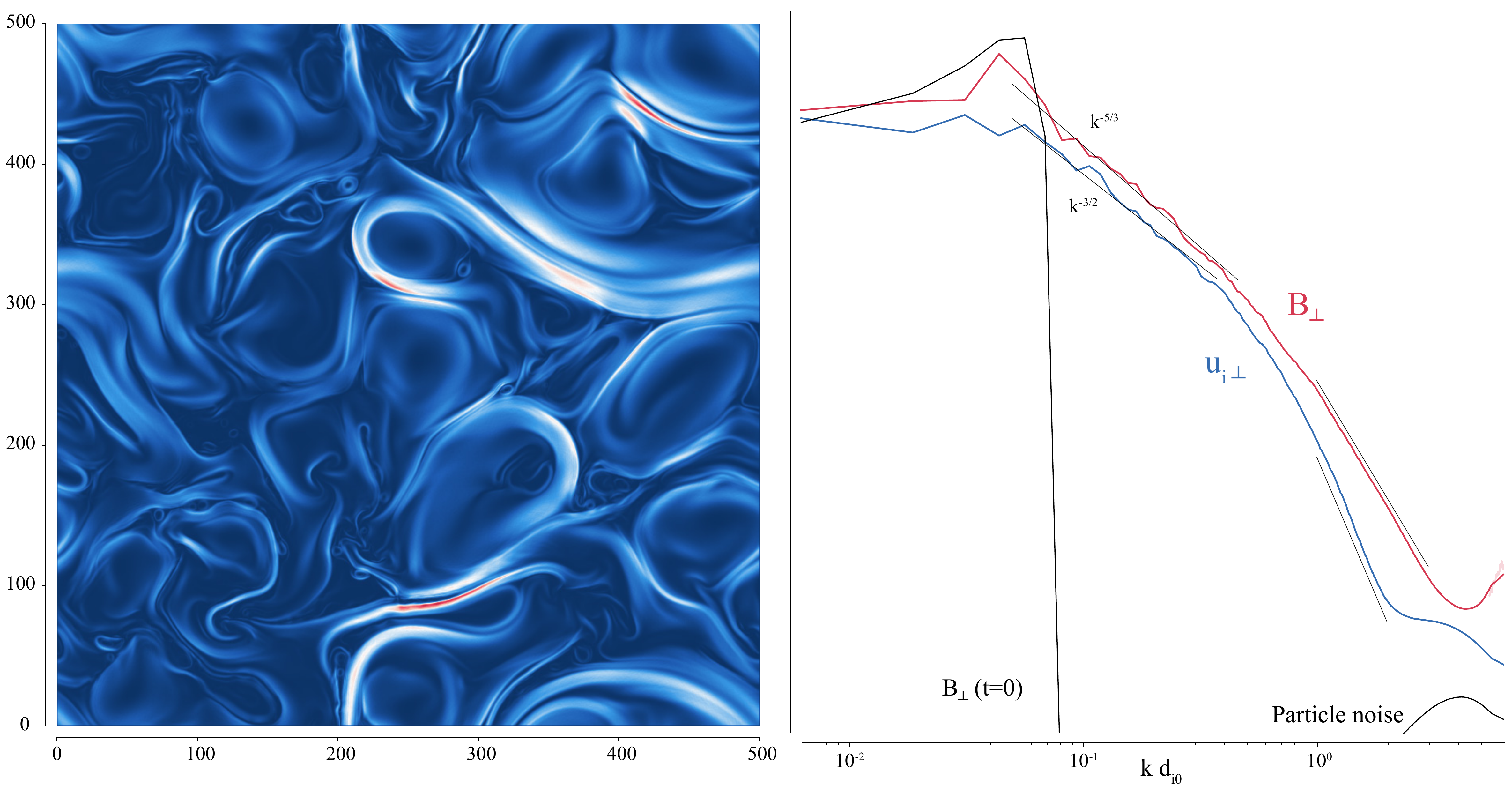}
        \caption{Decaying turbulence at time 500 $\omega_{ci0}^{-1}$.}
        \label{fig:decay}
    \end{figure}

    At time $t=500\ \omega_{ci0}^{-1}$, the perpendicular (in-plane) fluctuations of the magnetic field have reached the state displayed in Figure \ref{fig:decay}, left-hand panel. Vortices and current sheets give a maximum $B_{\perp}/B_0$ of about 1, a result consistent with solar wind turbulence observed at 1 au \cite{bruno2013turbulence}. The omni-directional power spectra of both the in-plane magnetic field fluctuations and the in-plane ion bulk velocity fluctuations are shown in the right-hand panel of the same figure. Omni-directional spectra are computed as follows, with $\hat f$ the (2D) Fourier transform of $f$:
    
    \begin{equation}
        P_{f}(k_\perp, k_\parallel) = |\hat f|^2
    \end{equation}
    
    These spectra are not further normalised and are given in arbitrary units. We then compute a binned statistics over this 2-dimensional array to sum up its values within the chosen bins of $k_\perp$, which correspond to rings in the $(k_\perp, k_\parallel)$-plane. The width of the rings is arbitrarily chosen so the resulting 1-dimensional spectrum is well resolved (not too few bins), and not too noisy (not too many bins). 
    
    \begin{equation}
        P_{f}(k_\perp) = \sum_{k_\perp \in [k_{\perp0},\ k_{\perp0}+\delta k_\perp]}|\hat f|^2
    \end{equation}
    
    For a vector field such as $\mathbf{B}_\perp=(B_x, B_y)$, the spectrum is computed as the sum of the spectra of each field component:
    
    \begin{equation}
        P_{\mathbf{B}_\perp}(k_\perp) = P_{B_x}(k_\perp) + P_{B_y}(k_\perp) .
    \end{equation}
    
    
    The perpendicular magnetic and kinetic energy spectra exhibit power laws over the inertial (MHD) range consistent with spectral indexes -5/3 and -3/2, respectively, between $k_\text{inj\, max}=0.1\ d_{i0}$ and break points around 0.5 $d_{i0}$. We remind that a spectral index -5/3 is consistent with the Goldreich-Sridhar strong turbulence phenomenology~\cite{Goldreich&Sridhar1997} that leads to a Kolmogorov-like scaling in the plane perpendicular to the background magnetic field, while a spectral index -3/2 is consistent with the Iroshnikov-Kraichnan scaling~\cite{Kraichnan1965PhFl}. These spectral sloped are themselves consistent with observations of magnetic and kinetic energy spectra associated to solar wind turbulence~\cite{Podestaetal2007,ChapmanHnat2007}. For higher wavenumbers, both spectral slopes get much steeper, and after a transition region within [0.5, 1.] $d_{i0}$ get to a value of about -3.2 and -4.5 when reaching the proton kinetic scales for respectively the perpendicular magnetic and kinetic energies, consistent with spectral index found at sub-ion scales by previous authors~\cite[e.g.]{franci2015apj, Sahraouietal2010}.
    Additionally, the initial spectra of the magnetic field and bulk velocity perturbations are over-plotted, to show respectively where the energy is injected in spatial frequencies and the level of noise introduced by the finite number of particles per node used.
    
    \begin{table}
        \centering
        \begin{tabular}{c|c}
            $B_0$ & 2.5 nT \\
            $n_0$ & 1 cm$^{-3}$ \\
            $\omega_{ci0}$ & 0.24 s \\
            $d_{i0}$ & 228 km \\
            $v_{A0}$ & 55 km/s \\
            $v_{th i0}$ & 55 km/s \\
            $\beta_{i0}=\beta_{e0}$ & 1 \\
            $B_{\perp 0}/B_0$ & 0.7 \\
        \end{tabular}
        \caption{Initial parameters of the decaying turbulence run}
        \label{tab:decay_param}
    \end{table}

\section{Step 2: Obstacle}
    \label{sec:step2}

    \emph{Menura} has shown satisfactory results on plasma turbulence, over three orders of magnitude in wavenumbers. We now start the second phase of the simulation, resuming it at $t=500\ \omega_{ci0}^{-1}$, corresponding to the snapshot studied in the previous section. We keep \emph{all} parameters unchanged, but add an obstacle with a relative velocity with regard to the frame used in the first phase, evolving through this developed turbulence. Particles and field are advanced with the exact same time and spatial resolutions as previously, so the interaction between this obstacle and the already-present turbulence is solved with the same self-consistency as in the first phase, with only one ingredient added: the obstacle.
    
    \subsection{A comet}
    
        This obstacle is chosen here to be an intermediate activity comet, meaning that its neutral outgassing rate is typical of an icy nucleus at a distance of about 2 au from the Sun. Comprehensive knowledge on this particular orbital phase of comets has recently been generated by the European \emph{Rosetta} mission, which orbited its host comet for two years \cite{glassmeier2007rosetta}. The first and foremost interest of such an object for this study is its size, which can be evaluated using the gyroradius of water ions in the solar wind at 2 au. The expected size of the interaction region is about 4 times this gyro-radius \cite{behar2018aa}, and with the characteristic physical parameters of Table~\ref{tab:decay_param}, the estimated size of the interaction region is 480 $d_{i0}$. In other words, the interaction region spans exactly over the range of spatial scales probed during the first phase of the simulation, including MHD and ion kinetic scales.
        
        The second interest of a comet is its relatively simple numerical implementation. Without a solid body, without gravity and without an intrinsic magnetic field, the obstacle is only made of cometary neutral particles being photo-ionised \emph{within} the solar wind. Over the scales of interest for this study, the neutral atmosphere can be modelled by a $1/r^2$ radial density profile, and considering the coma to be optically thin, ions are injected in the system with a rate following the same profile. This is the Haser Model \cite{haser1957distribution}, and simulating a comet over scales of hundreds of $d_{i0}$ only requires to inject cold cometary ions at each time step with the rate
        
        \begin{equation}\label{eq:haser}
        q_i(r) = \nu_i \cdot n_0(r) = \frac{\nu_i Q}{4 \pi u_0 r^2} ,
        \end{equation}
    
        with $r$ the distance from the comet nucleus of negligible size, $\nu_i$ the ionisation rate of cometary neutral molecules, $n_0$ the neutral cometary density, $Q$ the neutral outgassing rate, $u_0$ the radial expansion speed of the neutral atmosphere. 

    \subsection{Reference frame}
    
\begin{figure}
            \centering
            \includegraphics[width=\textwidth]{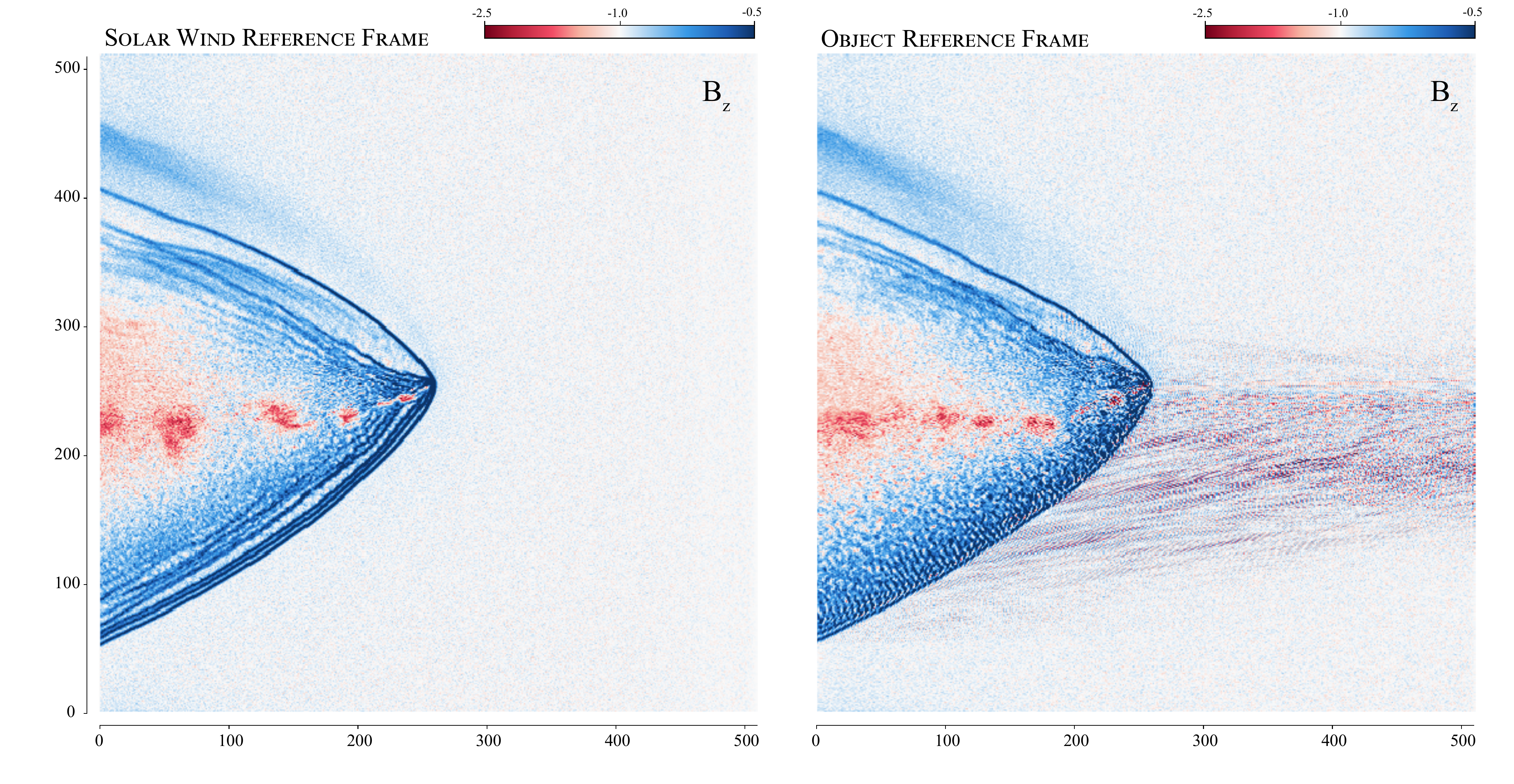}
            \caption{The interaction between a comet and a laminar flow, in the object rest frame (right) and the upstream solar wind reference frame (left). The magnetic field amplitude is shown. }
            \label{fig:orf_swrf}
        \end{figure}

        The first phase of the simulation, the decaying turbulence phase, was done in the plasma frame, in which the average ion bulk velocity is 0. Classically, planetary plasma simulation are done in the planet reference frame: the obstacle is static and the wind flows through the simulation domain. In this case, a global plasma reference frame is most of the time not defined. In \emph{Menura}, we have implemented the second phase of the simulation -- the interaction phase -- in the exact same frame as the first phase, which then corresponds to the plasma frame of the upstream solar wind, before interaction. In other words, the turbulent solar wind plasma is kept ``static'', and the obstacle is moving through this plasma. The reason motivating this choice is to keep the turbulent solar wind ``pristine'', by continuing its resolution over the exact same grid as in phase one. Another motivation for working in the solar wind reference frame is illustrated in Figure \ref{fig:orf_swrf}, in which we compare the exact same simulation done in each frame, using a laminar upstream flow. If the macroscopic result remains unchanged between the two frames, we find strong small scale numerical artifacts propagating upstream of the interaction in the comet reference frame, absent in the solar wind reference frame. Small scale oscillations are current in hybrid PIC simulations, and are usually filtered with either resistivity and/or hyper-resistivity, or with an ad-hoc smoothing method. Note that none of these methods are used in the present example. We demonstrate here the role of the reference frame in the production of one type of small scale oscillations, and insure that their influence over the spectral content of upstream turbulence is minimised, already without the implemented hyper-resistivity. 
    
        To summarise, by keeping the same reference frame during Step 1 and 2, the only effective difference between the two phases is the addition of sunward moving cometary macro-particle.
    
    \subsection{Algorithm}
    
\begin{figure}
            \centering
            \includegraphics[width=\textwidth]{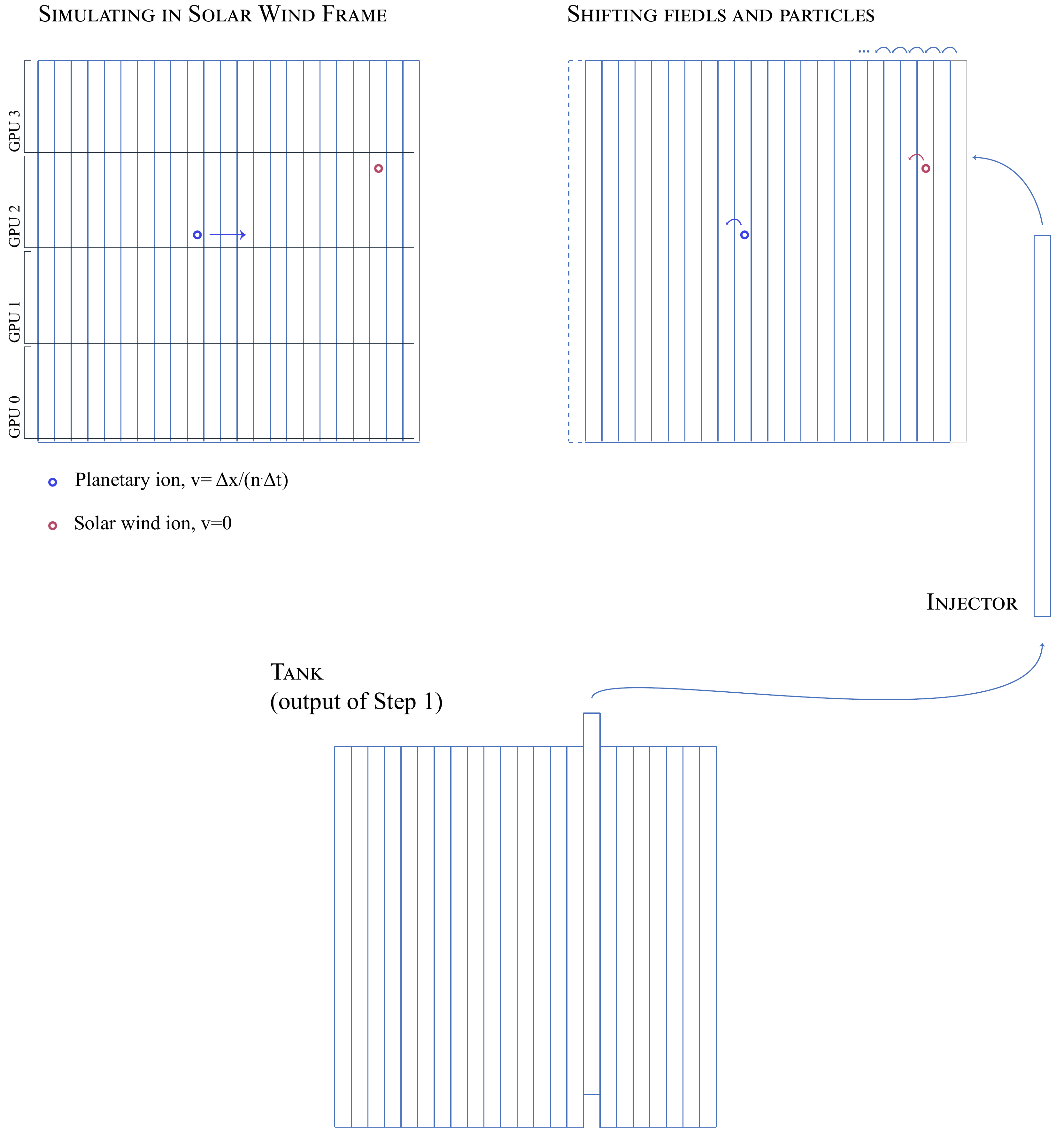}
            \caption{Injection algorithm for simulating a moving object within the simulation domain.}
            \label{fig:algo_inj}
        \end{figure}

        By working in the solar wind reference frame, the obstacle is moving within the simulation domain. Eventually, the obstacle would reach the boundaries of the box, before steady-state is reached. We therefore need to somehow keep the obstacle close to the centre of the simulation domain. This is done by shifting all particles and fields of $n \Delta x$ every $m$ iterations, $n, m \in \mathbb{N}$, as illustrated in Figure \ref{fig:algo_inj}. Using integers, the shift of the field is simply a side-way copy of themselves without the need of any interpolation, and the shift of the particles is simply the subtraction of $n \Delta x$ to their $x$-coordinate. Field values as well are particles ending up downstream of the simulation domain are discarded. \\
        This leaves only the injection boundary to be dealt with. There, we simply inject a slice of fields and particles picked from the output of Step 1, using the right slice index in order to inject the continuous turbulent solution, as shown in Figure \ref{fig:algo_inj}. These slices are $n \Delta x$ wide.
        
        With \texttt{idx\_it} the index of the iteration, the algorithm illustrated in Figure \ref{fig:algo_inj} is then:
        
        \begin{itemize}
            \item Inject cometary ions according to $q_i(r)$ (cf. Eq. \ref{eq:haser})
            \item Advance particles and fields (cf. Figure \ref{fig:algo})
            \item If \texttt{idx\_it\%m=0}
                \begin{itemize}
                    \item Shift particles and fields of \texttt{-n$\Delta$x}
                    \item Discard downstream values
                    \item Inject upstream slice \texttt{idx\_{slice}} from Step 1 output
                    \item Increment \texttt{idx\_{slice}}
                \end{itemize} 
            \item Increment \texttt{idx\_it}
        \end{itemize}
    
        This approach has one constraint, we cannot fine-tune the relative speed $v_0$ between the wind and the obstacle, which has to be
        
        \begin{equation}
            v_0 = \frac{n}{m} \frac{\Delta x}{\Delta t}
        \end{equation}
        
        in order for the obstacle to come back to its position every \texttt{m} iterations, and therefore not drift up- or downstream of the simulation domain.
    
    \subsection{CUDA and MPI implementation, performances}
    
        The computation done by \emph{Menura}'s solver (Figure \ref{fig:algo}) is entirely executed on multiple GPUs (Graphics Processing Units), written in \texttt{c++} in conjunction with the \texttt{CUDA} programming model and the \texttt{MPI} standard, which allows to split the problem and distribute it over multiple cards (i.e. processors). GPUs can run simultaneously thousands of threads, and can therefore tremendously accelerate such applications. The first implementation of a hybrid-PIC model on such devices was done by \cite{fatemi2017jp}. However their still limited memory (up to 80GB at the time of writing) is a clear constraint for large problems, especially for turbulence simulation which requires a large range of spatial scales \emph{and} a very large number of particles per grid node. The use of multiple cards becomes then unavoidable, and the communications between them is implemented using a CUDA-aware version of MPI, the Message Parsing Interface. The division of the simulation domain in the current version of \emph{Menura} is kept very simple, with equal size rectangular sub-domains distributed along the direction perpendicular to the motion of the obstacle: one sub-domain spans the entire domain along the x-axis with its major dimension, as shown in Figure \ref{fig:algo_inj}. MPI communications are done for particles after each position advancement, and for fields after each solution of the Ohm's law and the Faraday's law. But since the shift of fields and particles described in the previous section is done purely along the obstacle motion direction, no MPI communication is needed after the shifts, thanks to the orientation of the sub-domains.
        
        Another limitation in using GPUs is the data transfer time between the CPU and the cards. In \emph{Menura}, all variables are initialised on the CPU, and are saved from the CPU. Data transfers are then unavoidable, before starting the main loop, and every time we want to save the current state of the variables. During Step 2 of the simulation, a copy of the outputs of Step 1 is needed, which effectively doubles the memory needed for Step 2. This copy is kept on the CPU (in the \texttt{tank} object) in order to make the most out of the GPUs memory, in turn implying that more CPU-GPU communications are needed for this second step. Every time we inject a slice of fields and particles upstream of the domain, only this amount of data is copied from the CPU to the GPUs, using the \texttt{injector} variable as sketched in Figure \ref{fig:algo_inj}.\\

        For Step 1, the decaying turbulence ran 10000 iterations, four nvidia V100 GPUs were used with 16GB memory each, corresponding to one complete node of the IDRIS cluster Jean Zay. A total of 2 billion particles (500 million per card) were initialised. The time for the solver on each card reached a bit more than three hours, with a final total coast of about 13 hours of computation time for this simulation, taking into account all four cards, and the variables initialisation and output. Step 2 was executed on larger V100 32GB cards, providing much more room for the addition of 60000 cometary macro-particles per iteration.
        
        During Step 1, 87.3\% of the computation time was spent on moments mapping, i.e. steps 0 and 2 in the algorithm of Figure \ref{fig:algo}, while respectively 2.7\% and 0.8\% were spent on advancing the particles velocity and position. The computation of the Ohm's and Faraday's laws sums up to 0.5\%. 0.9\% was utilised for MPI communications of field variables, while only 0.08\% was dedicated for particles MPI communications, due to the limited particle transport happening in Step 1.
        
        91\% of the total solver computation time is devoted to particles treatment, with 96\% of that part spent on particles moment mapping, which might seem a suspiciously large fraction. We note however that such a simulation is characterised by its large number of particles per node, 2000 in our case. 99.6\% of the total allocated memory is devoted to particles. The time spent to map the particles on the grid is also remarkably larger than the time spent to update their velocity, despite both operations being based on the same interpolation scheme. However, during the mapping of the particles moments, thousands of particles need to \emph{increment} the value at particular memory addresses (corresponding to macro-particle density and flux), whereas during the particle velocity advancement, thousands of particles only need to \emph{read} the value of the same addresses (electric and magnetic field).
        


\section{First result}
    \label{sec:result}

\begin{figure}
        \centering
        \includegraphics[width=1.\textwidth]{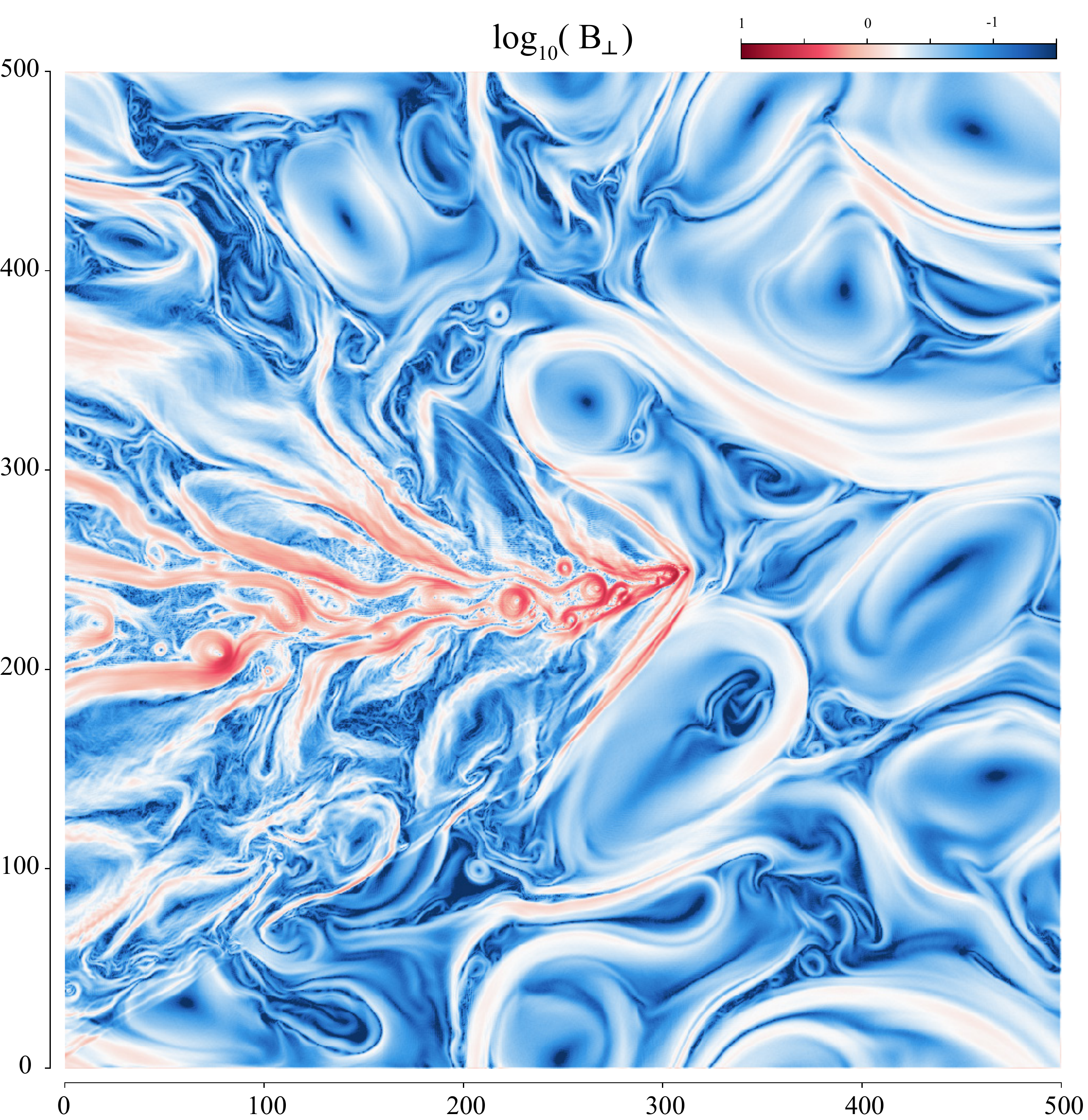}
        \caption{Perpendicular magnetic fluctuations during the interaction.}
        \label{fig:result}
    \end{figure}

    We now focus on the result of Step 2, in which cometary ions were steadily added to the turbulent plasma of Step 1, moving at a super-Alfv\'enic and super-sonic speed. Table \ref{tab:comet_param} lists the physical parameters used for Step 2. After 4000 iterations, the total number of cometary macro-particles in the simulation domain reaches a constant average value: the comet is fully developed and has reached an average ``steady'' state. From this point, we simulate more than one full injection period, 1500 iterations after which we loop over the domain of the injection tank in Figure \ref{fig:algo_inj}. As an example, Figure \ref{fig:result} displays the state of the system at iteration 6000, focusing here again on the perpendicular fluctuations of the magnetic field. This time the colour scale is logarithmic, since magnetic field fluctuations are spanning over a much wider range than previously. While being advected through a dense cometary atmosphere, the solar wind magnetic field \emph{piles up} (augmentation of its amplitude because of the slowing down of the total plasma bulk velocity) and \emph{drapes} (deformation of its field lines due to the differential pile-up around the density profile of the coma), as first theorised by \cite{alfven1957theory}. This general result was always applied to the global, average magnetic field, and was observed in situ at the various comets visited by space probes.
    
    Without diving very deep in the first results of \emph{Menura}, we see that the pile-up and the draping of upstream perpendicular magnetic field fluctuations also has an important impact on the tail of the comet, with the creation of large amplitude magnetic field vortices of medium and small size. This phenomenon, together with a deeper exploration of the impact of solar wind turbulence on the physics of a comet, are gathered in a subsequent publication. 
    
    \begin{table}[]
        \centering
        \begin{tabular}{c|c}
            $u_0$ & 363 km/s (6.66 $v_A$) \\
            $Q$ & $5\cdot 10^{26}$ s$^{-1}$\\
            $\nu_i$ & $2\cdot 10^{-7}$ s$^{-1}$\\
            $u_0$ & 1 km/s\\
        \end{tabular}
        \caption{Physical parameters of the comet.}
        \label{tab:comet_param}
    \end{table}

\section{Conclusion}


   This publication introduces a new hybrid PIC plasma solver, \emph{Menura}, that allows for the first time to investigate the impact of a turbulent plasma flow on an obstacle. For this purpose, a new 2-step simulation approach has been developed which consist in, first, developing a turbulent plasma, and second, injecting it periodically in a box containing an obstacle. The model has been validated with respect to well-known fluid and kinetic plasma phenomena. \emph{Menura} has also proven to provide the results expected at the output of this first step of the model -- namely decaying magnetised plasma turbulence.  
    
    Until now, all planetary science oriented simulations have ignored all-together the turbulent nature of the solar wind plasma flow, in terms of structures and in terms of energy. \emph{Menura} has been design to fulfill this lack and it will now allow us to explore, for the first time, some fundamental questions that have remained open regarding the impact of the solar wind on different solar system objects, such as: what happens to turbulent magnetic field structures when it impacts on an obstacle such as a magnetosphere? How are the properties of a turbulent plasma flow reset as is crosses a shock, such as the solar wind crossing a planetary bow shock? How do the additional energy, stored in the perpendicular magnetic and velocity field components impact the large-scale structures and dynamics of planetary magnetospheres?  
    
    On top of the study of the interaction between the turbulent solar wind and solar system obstacles, we are confident that the new modeling framework developed in this work, in particular its 2-step approach might as well be suitable for the study of energetic solar wind phenomenons, namely Co-rotating Interaction Regions and Coronal Mass Ejections, which could be similarly simulated first in the absence of an obstacle, to then be used as inputs of a second step including obstacles.

\section{Acknowledgements}

E. Behar acknowledges support from Swedish National Research Council, Grant 2019-06289. This research was conducted using computational resources provided by the Swedish National Infrastructure for Computing (SNIC), Project SNIC 2020/5-290 and SNIC 2021/22-24 at the High Performance Computing Center North (HPC2N), Umeå University, Sweden. This work was granted access to the HPC resources of IDRIS under the allocation 2021-AP010412309 made by GENCI. Work at LPC2E and Lagrange was partly funded by CNES.

\appendix

\section{Normalised equations}
\label{app:normalised_equations}

    In \emph{Menura}'s solver, all variables are normalised using the background magnetic field amplitude $B_0$ and background density $n_0$, or equivalently using the corresponding proton gyrofrequency $\omega_{ci0}$ and Alfv\'en speed  $v_{A0}$. We report the table of the background variables definitions in Table \ref{tab:normalisation2}.

    \begin{table}[]
        \centering
        \begin{tabular}{c|c}
            $B_0$ & $B_0$ \\
            $n_0$ & $n_0$ \\
            $v_0$ & $v_{A0}=B_0/\sqrt{\mu_0 m_i n_0}$\\
            $\omega_0$ & $\omega_{ci0}=e B_0/m_i$\\
            $x_0$ & $d_{i0}=v_{A0}/\omega_{ci0}$ \\
            $t_0$ & $1/\omega_{ci0}$\\
            $E_0$ & $v_{A0}\cdot B_0$\\
            $p_0$ & $B_0^2/(2\mu_0)$\\
            $m_0$ & $m_i n_0 x_0^3$\\
            $q_0$ & $e\, n_0 x_0^3$\\
        \end{tabular}
        \caption{Caption}
        \label{tab:normalisation2}
    \end{table}

    Based on these definitions, one can derive the following main equations of the solver. A normalised variable $\tilde{a}$ is obtained by dividing this variable by its background value, $\tilde{a} = a/a_0$, equivalently $a = \tilde{a}\, a_0$. We first consider the Faraday's law, and using the background parameters definitions of Table \ref{tab:normalisation2}:

    \begin{align}
        \frac{\partial \mathbf{B}}{\partial t} & = -\mathbf{\nabla} \times \mathbf{E} \\
        \Rightarrow \frac{\partial \tilde{\mathbf{B}} B_0}{\partial \tilde{t}\, t_0} & = -\tilde{\mathbf{\nabla}}/x_0 \times (\tilde{\mathbf{E}}\, E_0) \\
        \Rightarrow \frac{\partial \tilde{\mathbf{B}} B_0}{\partial \tilde{t} /\omega_{ci0}} & = -\tilde{\mathbf{\nabla}}/d_{i0} \times (\tilde{\mathbf{E}}\ v_{A0}\, B_0) \\
        \Rightarrow \frac{\partial \tilde{\mathbf{B}}}{\partial \tilde{t}} & = -\tilde{\mathbf{\nabla}} \times \tilde{\mathbf{E}}
    \end{align}

    with
    
    \begin{equation}
        \tilde{\mathbf{\nabla}} = \left(\frac{\partial}{\partial \tilde{x}}, \ \frac{\partial}{\partial \tilde{y}}\right)
    \end{equation}
    In other words, the Faraday's law expressed with normalised variables is unchanged compared to its SI definition. The Ohm's law becomes:

    \begin{equation}
        \tilde{\mathbf{E}} = -\tilde{\mathbf{u}_i} \times \tilde{\mathbf{B}} + \tilde{\mathbf{J}} \times \tilde{\mathbf{B}} +  \tilde{\mathbf{\nabla}}\cdot \tilde{p_e} - \tilde{\eta_h} \tilde{\mathbf{\nabla}}^2 \tilde{\mathbf{J}}
    \end{equation}
    
    with
    
    \begin{equation}
        \tilde{\mathbf{J}} = \tilde{\mathbf{\nabla}} \times \tilde{\mathbf{B}}
    \end{equation}

    and

    \begin{equation}
        \tilde{p_e} = \beta_e\ \tilde{n}^\kappa
    \end{equation}

    Concerning the gathering of particles moments,

    \begin{equation}
        \tilde{n} = \sum_\text{spec} w_\text{spec} \sum_p W(\tilde{\mathbf{r}_p})
    \end{equation}
    
    with $w_\text{spec} = \tilde{n}_\text{spec}/ \text{particle-per-node}_\text{spec}$ . For the solar wind proton, $\tilde{n}=1$ and one simply gets $w_\text{sw} = 1/ \text{particle-per-node}$ . $W(\tilde{\mathbf{r}_p})$ stands for the shape factor, triangular in our case (in 2 spatial dimensions, one macro-particle affects the density and current of nine grid nodes, with linear weights).
    
    \begin{equation}
        \tilde{J}_i = \sum_\text{spec} w_\text{spec} \sum_p \tilde{\mathbf{u}_i}(\tilde{\mathbf{r}_p}) W(\tilde{\mathbf{r}_p})
    \end{equation}


\bibliographystyle{elsarticle-num} 
\bibliography{cas-refs}





\end{document}